\documentclass[12pt]{article}
\usepackage{amssymb,latexsym}
\oddsidemargin=\evensidemargin
\advance\topmargin by-\headheight
\def\aut{\mathop{\rm Aut}\nolimits}

\def\G{{\cal G}}
\def\GCD{\mathop{\rm GCD}\nolimits}

\def\im{\mathop{\rm im}\nolimits}
\def\ker{\mathop{\rm ker}\nolimits}
\def\lg{\langle}
\def\mapright#1{\smash{\mathop{\rightarrow}\limits^{#1}}}
\def\mod{\mathop{\rm mod}\nolimits}
\def\NN{{\mathbb N}}
\def\ov{\overline}
\def\P{{\cal P }}
\def\R{{\cal R}}
\def\rg{\rangle}
\def\S{{\cal S}}
\def\sym{\mathop{\rm Sym}\nolimits}
\def\T{{\cal T}}
\def\wt{\widetilde}
\def\wk{\widetilde k}
\def\wH{\widetilde H}
\def\wK{\widetilde K}

\def\ZZ{{\mathbb Z}}

\def\proof{{\bf Proof}.\ }
\def\bull{\vrule height .9ex width .8ex depth -.1ex }
\def\subs{\stepcounter{subsection}{\bf\thesubsection{.}}
\addtocounter{subsection}{-1}\refstepcounter{subsection}}
\def\subsn{\vspace{2mm}\subs}

\newtheorem{formula}{}[section]

\newtheorem{definition}[formula]{Definition}
\newtheorem{corollary}[formula]{Corollary}
\newtheorem{remark}[formula]{Remark}
\newtheorem{lemma}[formula]{Lemma}
\newtheorem{theorem}[formula]{Theorem}

\def\thrm{\begin{theorem}}
\def\thrml#1{\begin{theorem}\label{#1}}
\def\ethrm{\end{theorem}}
\def\rmrk{\begin{remark}}
\def\rmrkl#1{\begin{remark}\label{#1}}
\def\ermrk{\end{remark}}
\def\dfntn{\begin{definition}}
\def\dfntnl#1{\begin{definition}\label{#1}}
\def\edfntn{\end{definition}}
\def\nmrt{\begin{enumerate}}
\def\enmrt{\end{enumerate}}
\def\tm#1{\item[{\rm (#1)}]}
\def\qtn{\begin{equation}}
\def\qtnl#1{\begin{equation}\label{#1}}
\def\eqtn{\end{equation}}
\def\lmm{\begin{lemma}}
\def\lmml#1{\begin{lemma}\label{#1}}
\def\elmm{\end{lemma}}
\def\crllr{\begin{corollary}}
\def\crllrl#1{\begin{corollary}\label{#1}}
\def\ecrllr{\end{corollary}}

\begin{document}
\title{On non-abelian homomorphic public-key cryptosystems}
\author{
Dima Grigoriev \\[-1pt]
\small IRMAR, Universit\'e de Rennes \\[-3pt]
\small Beaulieu, 35042, Rennes, France\\[-3pt]
{\tt \small dima@maths.univ-rennes1.fr}\\[-3pt]
\small http://www.maths.univ-rennes1.fr/\~{}dima
\and
Ilia Ponomarenko\\[-1pt]
\small Steklov Institute of Mathematics,\\[-3pt]
\small Fontanka 27, St. Petersburg 191011, Russia\\[-3pt]
{\tt \small inp@pdmi.ras.ru}
}
\date{12.11.2002}
\maketitle

\begin{abstract}
An important problem of modern cryptography concerns
secret public-key computations in algebraic structures.
We construct homomorphic cryptosystems
being (secret) epimorphisms $f:G\to H$, where $G,H$ are (publically known)
groups and $H$ is finite. A letter of a message to be encrypted is an
element $h\in H$, while its encryption $g\in G$ is such that
$f(g)=h$. A homomorphic cryptosystem allows one to perform
computations (operating in a group $G$) with encrypted
information (without knowing the original message over $H$).

In this paper certain homomorphic cryptosystems are constructed for
the first time for non-abelian groups $H$ (earlier, homomorphic
cryptosystems were known only in the Abelian case). In fact, we
present such a system for any solvable (fixed) group~$H$.
\end{abstract}

\section{Introduction}

In what follows all the groups are presented in  some natural
way depending on the problem. For example, the special
constructions of Section~\ref{expc} are based on
the groups $\ZZ_n^+$ and $\ZZ^*_n$ just given via
$n$, whereas the general construction of
Section~\ref{frpr} requires only that elements of a group
in question can be generated and moreover, the multiplication and
taking the inverse in the group can be performed efficiently. In the latter
case the groups can be presented by generators and relations
or even by generic algorithms (see e.g.~\cite{MW}).

There is a lot of public-key cryptosystems using groups (see e.g.
\cite{B,KL,KMOV,NS,Ra,Ri}) but only a few of them have a homomorphic property
in the sense of the following definition (cf. also~\cite{FM,SYY,Y}).

\dfntnl{md}
Let $H$ be a finite non-identity group, $G$ a finitely generated group
and $f:G\to H$ an epimorphism. Suppose that $R$ is a set of distinct
representatives of the right cosets of $G$ with respect to $\ker(f)$,
$A$ is a set of words in some alphabet and a mapping $P:A\to G$ such that
$\im(P)=\ker(f)$. A triple $\S=(R,A,P)$ is called a {\it homomorphic
cryptosystem} over $H$ with respect to $f$, if the following conditions
are satisfied:
\nmrt
\tm{H1} one can get random elements (of the sets $A,G,H$), compute the inverse
of an element and the product of two elements (in the group $G$ or $H$) in
polynomial in $N$ probabilistic time where $N$ is the size of presentations
of $G,H$ and $A$;
\tm{H2} $|R|=|H|$ and for any element $g\in R$ its image $f(g)$ as well as for
any element $h\in H$ its unique preimage $g\in R$ such that $f(g)=h$ can be
computed in polynomial in $N$ probabilistic time;
\tm{H3} the mapping $P$ is a trapdoor function.
\enmrt
\edfntn

\rmrk
We require that the set $R$ is given explicitly by a list of elements of
$G$. So, condition (H2) implies that without loss of generality one can assume
that the group $H$ is represented by its multiplication table.
\ermrk

Condition (H3) means (see~\cite{GB}) that the values of $P$ can
be computed in polynomial in $N$  probabilistic  time, whereas
finding of the inverse mapping $P^{-1}$ is a hard computational problem
which can be solved with the help of some additional secret information (for
instance, knowing some invariant of the group $G$). In a homomorphic
cryptosystem $\S$ the elements of $H$ are (publically) encrypted in a probabilistic manner by the elements of $G$,
all the computations are performed in $G$ and the result is decrypted to $H$.
More precisely:

\vspace{4mm}
\noindent{\bf Public Key:} $G,H,R,A,P,f|_R$.

\vspace{2mm}
\noindent{\bf Secret Key:} finding $P^{-1}$.

\vspace{2mm}
\noindent{\bf Encryption:} given a plaintext $h\in H$ take $r\in R$ such
that $f(r)=h$ (invoking (H2)) and a random element $a\in A$; the ciphertext
of $h$ is the element $P(a)r$ of~$G$ (the element $a$ as well as the product
$P(a)r$ is computed by means of (H1)).

\vspace{2mm}
\noindent{\bf Decryption:} given $g\in G$ find the elements $r\in R$
and $a\in A$ such that $rg^{-1}=P(a)$ (for computing $P(a)$ see (H3)); set
the plaintext of $g$ to be $f(g)=f(r)$ (the element $f(r)$ is computed
by means of (H2)).
\vspace{4mm}

One can see that the encryption procedure can be performed
by means of public keys efficiently. However, the decryption procedure
is a secret one in the following sense. To find the element $r$ one
has to solve in fact, the membership problem for the subgroup $\ker(f)$
of the group~$G$. We assume that a solution for each instance $g'\in\ker(f)$
of this problem must have a ``proof'', which is actually an element
$a\in P^{-1}(g')$. Thus, the secrecy of the system is based on the assumption
that finding an element in the set $P^{-1}(g')$ is an intractable computation
problem. On the other hand, our ability to compute $P^{-1}$ enables us to
efficiently implement the decryption algorithm. One can treat $P$ as a proof
system for $\ker(f)$ in the sense of~\cite{CR}. Moreover, in case when $A$
is a certain group and $P$ is a homomorphism we have the following {\it exact}
sequence of group homomorphisms
$$
A\,\mapright{P}\,G\,\mapright{f}\,H\,\mapright{}\,\{1\}
$$
(recall that the exact sequence means that the image of each homomorphism in
it coincides with the kernel of the next one).

In the present paper the group $H$ being an alphabet of plaintext messages
is always finite (and rather small) and given by its multiplication table,
while the group $G$ of ciphertext messages could be infinite but being always
finitely generated. However, the infinitness of~$G$ is not an obstacle for encrypting
(and decrypting) since an element from $H$ is encrypted by a finite word in
generators of $G$. For example, in \cite{DJSS} for an (infinite, non-abelian in
general) group $H$ given by $m$ generators and relations a natural epimorphism
$f:F_m \rightarrow H$ from a free group $F_m$ is considered. Thus, for any
element of $H$ one can produce its preimages (encryptions) by inserting in a
word (being already a produced preimage of $f$) from $F_m$ any relation
defining $H$. In other terms, decrypting of $f$ reduces to the word problem
in $H$. The main difference with our approach is that we consider free
products over groups of number-theoretic nature like $Z_n^*$ (rather than
given by generators and relations). This allows one to provide evidence for
difficulty of decryption.

\dfntnl{gc}
$\G_{crypt}$ is the class of all finite groups $H$ for which there exists a
homomorphic cryptosystem over $H$.
\edfntn

In the context of our definition of a homomorphic cryptosystem the main
problem we study in this paper is to prove that the class $\G_{crypt}$
contains all solvable nonidentity groups (see Theorem~\ref{th3}).

To our knowledge all known at present homomorphic cryptosystems
are more or less modifications of the following one. Let $n$ be the product
of two distinct large primes of size $O(\log n)$. Set
$G=\{g\in \ZZ_n^*:\ {\bf J}_n(g)=1\}$ where ${\bf J}_n$ is the Jacobi symbol,
and $H=\ZZ_2^+$. Then given a non-square $g_0 \in G$ the triple $(R,A,P)$
where
$$
R=\{1,g_0\},\quad A=\ZZ_n^*,\quad P(g):g\mapsto g^2,
$$
is a homomorphic cryptosystem over $H$ with respect to the natural epimorphism
$f:G\to H$ with $\ker(f)=\{g^2:\ g\in\ZZ_n^*\}$ (see~\cite{GM,GB}). We call it the
{\it quadratic residue cryptosystem}. It can be proved (see \cite{GM,GB})
that in this case finding $P^{-1}$ is not easier than factoring $n$, whereas
given a prime divisor of $n$ the computation of $P^{-1}$ can be performed in
polynomial time in $\log n$.

It is an essential assumption (being a shortcoming) in the quadratic residue
cryptosystem as well as other cryptosystems cited below that its security
relies on a fixed a priori (proof system) $P$. Indeed, it is not excluded
that adversary could verify whether an element of $G$ belongs to $\ker (f)$
avoiding making use of~$P$, for example, in case of the quadratic residue
cryptosystem that would mean verifying that $g\in G$ is a square without
providing a square root of~$g$. Although, there is a common conjecture that
verifying for an element to be a square (as well as some power) is also
difficult.

Let us mention that a cryptosystem from \cite{P99} over $H=\ZZ_n^+$ (for the
same assumptions on $n$ as in the quadratic residue cryptosystem) with respect
to the homomorphism $f:G\to H$ where $G=\ZZ_{n^2}^*$ and
$\ker(f)=\{g^n:\ g\in G\}$, in which $A=G$ and $P:g\mapsto g^n$, is not
homomorphic in the sense of Definition~\ref{md} because condition (H3) of it
does not hold. (Since $|G|\le |H|^2$, one can inverse $P$ in a polynomial
time in $|H|$.) By the same reason the cryptosystem
from \cite{OU98} over $H=\ZZ_{p}^+$ with respect to the homomorphism
$f:G\to H$ where $G=\ZZ_{p^2q}^*$ and $\ker(f)=\{g^{pq}:\ g\in G\}$
(here the integers $p,q$ are distinct large primes of the same size) is also
not homomorphic (besides, in this system only a part of the group $H$ is
encrypted). Some cryptosystems over certain dihedral groups were studied
in~\cite{Ra}.

We note in addition that an alternative setting of a homomorphic (in fact,
isomorphic) encryption $E$ (and a decryption $D=E^{-1}$) was proposed
in~\cite{KMOV}. Unlike Definition~\ref{md} the encryption $E:G\rightarrow G$
is executed in the same set $G$ (being an elliptic curve over the ring~$\ZZ_n$)
treated as the set of plaintext messages. If $n$ is composite, then $G$ is not
a group while being endowed with a partially defined binary operation which
converts $G$ in a group when $n$ is prime. The problem of decrypting this
cryptosystem is  close to the factoring of~$n$. In
this aspect \cite{KMOV} is similar to the well-known RSA scheme (see e.g.
\cite{GB}) if to interprete RSA as a homomorphism (in fact, isomorphism)
$E:Z_n^*\rightarrow Z_n^*$, for which the security relies on the difficulty
of finding the order of the group $Z_n^*$.

We complete the introduction by mentioning some cryptosystems using groups
but not being homomorphic in the sense of Definition~\ref{md}. The well-known
example is a cryptosystem which relies on the Diffie-Hellman key agreement
protocol (see e.g.~\cite{GB}).
It involves cyclic groups and relates to the discrete logarithm
problem~\cite{MW}; the complexity of this system was studied in~\cite{Sp}.
Some generalizations of this system to non-abelian groups (in particular,
the matrix groups over some rings) were suggested in~\cite{PKHK} where
secrecy was based on an analog of the discrete logarithm problems in groups
of inner automorphisms. Certain variations of the Diffie-Hellman systems over the
braid groups were described in~\cite{KL}; here several trapdoor one-way
functions connected with the conjugacy and the taking root problems in the
braid groups were proposed. Finally it should be noted that a cryptosystem
from~\cite{NS} is based on a monomorphism $\ZZ_m^+\to\ZZ_n^*$ by
means of which $x$ is encrypted by $g^x\,(\mod n)$ where $n,g$ constitute
a public key; its decrypting relates to the discrete logarithm problem and is
feasible in this situation due to a special choice of $n$ and $m$ (cf.
also~\cite{B}).

\section{Homomorphic cryptosystems over cyclic groups}
\label{expc}

In this section we present an explicit homomorphic cryptosystem over a
cyclic group of a prime order $m$ whose decription is based on taking
$m$-roots in the group $\ZZ^*_n$ for a suitable $n\in\NN$. It can be
considered in a sense as a generalization
of the quadratic residue cryptosystem over $\ZZ_2^+$.  Throughout
this section given $n\in\NN$ we denote by $|n|$ the size of the number~$n$.

Given $m,N\in\NN$ set
$\T_N=\{(p,q):\ p,q\ \textstyle{\rm are\ primes},\ |p|=|q|=N,\ p<q\}$
and
$$
D_{N,m}=\{n\in\NN:\ n=pq,\ (p,q)\in\T_N,\ m|p-1,\ \GCD(m,q-1)=1\}.
$$
From the Dirichlet's theorem on primes in arithmetic progressions \cite{D} it
follows that given an odd prime~$m$, the set $D_{N,m}$ is not empty for
sufficiently large numbers~$N$.

Let $n\in D_{N,m}$ for some natural number $N$ and an odd prime~$m$. Then the
group $G=\ZZ_n^*$ has a (normal) subgroup $G_0=\{g^m:\ g\in G\}$ the factor
by which is isomorphic to the group $H=\ZZ_m^+$. Denote by $f$ the
corresponding epimorphism from $G$ to $H$. The mapping
\qtnl{l299}
P:G\to G,\quad g\mapsto g^m
\eqtn
is obviously a polynomial time computable homomorphism such that
$\im(P)=\ker(f)$. Next, any element of the set
$$
R_{m,n}=\{R\subset G:\ |f(R)|=|R|=m\}
$$
is a system of distinct representatives of the cosets of $G$ by $G_0$. We
observe that given the decomposition $n=pq$ one can find an element
$R\in R_{m,n}$ in probabilistic time $|n|^{O(1)}$. Indeed, since $m$ is a
prime, it suffices to compute a random element $s_p\in\ZZ_p^*$ such that
$s_p^{(p-1)/m}\ne 1$, and an element $s_q\in\ZZ_q^*$, then find by the
Chineese reminder theorem the unique element $s\in\ZZ_q^*$ such that
$s=s_p\,(\mod p)$, $s=s_q\,(\mod q)$, and set
$R=\{s^it_i^m:\ i=0,\ldots,m-1\}$ for arbitrary elements $t_i\in\ZZ_n^*$.

We claim that the triple $\S_{N,m,n}=(R,A,P)$ with arbitrary chosen set
$R\in R_{m,n}$, $A=G$ and $P$ defined by~(\ref{l299}) is a homomorphic
cryptosystem over the group $H$ with respect to the epimorphism~$f$ whenever
the following statement is true:

\vspace{4mm}
\noindent{\bf Assumption (*).} For an odd prime $m$ the problem $\P(m)$, of
finding the $m$-root in $\ZZ_n^*$ with $n\in D_{N,m}$ given an element
$R\in R_{m,n}$ is not easier than the same problem without any such $R$.
\vspace{4mm}

Let us present the group $G$ by the number $n$ and the group $H$ by the
set of its elements. Then for the triple $S_{N,m,n}$ conditions (H1) and (H2)
of Definition~\ref{md} are trivially satisfied (the image of the above
element $s^it^m_i$ with respect to the homomorphism~$f$ equals $i\in\ZZ^+_m$).
In fact, condition (H3) would follow from the next lemma.

\lmml{l7}
Let $N\in\NN$, $m$ be an odd prime and $n\in D_{N,m}$. Then
\nmrt
\tm{1} given primes $p$ and $q$ such that $n=pq$ and an element $g\in\ZZ_n^*$
one can verify whether $g$ is an $m$-power and if it is the case one can find
an $m$-root of $g$ in probabilistic polynomial time in $N$;
\tm{2} the factoring problem for $n$ is probabilistic polynomial time
reducible to the problem of finding an $m$-root in $\ZZ_n^*$.
\enmrt
\elmm
\proof Throughout the proof we will use the canonical decomposition
$\ZZ_n^*=\ZZ_p^*\times\ZZ_q^*$. To prove statement (1) we make use of Rabin's probabilistic
polynomial-time algorithm for finding roots of polynomials over finite prime
fields (see~\cite{R}). Namely, given the primes $p,q$ and $g\in\ZZ_n^*$ we
proceed as follows:

\begin{itemize}
\item[]{\bf Step 1.} Find the elements $g_p\in\ZZ_p^*$ and $g_q\in\ZZ_q^*$
such that $g=g_p\times g_q$, i.e. $g_p=g\ \,(\mod p)$, $g_q=g\ \,(\mod q)$.
\vspace{2mm}

\item[]{\bf Step 2.} By Rabin's algorithm (for a prime field) find some roots
$h_p\in\ZZ^*_p$ and $h_q\in\ZZ^*_q$ of the polynomials $x^m-g_p$ and
$x^m-g_q$, respectively.
\vspace{2mm}

\item[]{\bf Step 3.} Output $h=h_p\times h_q$.
\end{itemize}
\vspace{2mm}

\noindent Observe that the described algorithm fails (at Step~2) if and only
if $g$ is not an $m$-power. Since, obviously, $h^m=h_p^m\times h_q^m=
g_p\times g_q=g$, statement (1) of the lemma is proved.

To prove statement (2) suppose that we are supplied  with a
probabilistic polynomial-time algorithm $Q_n$ that given $g\in\ZZ_n^*$
computes an $m$-root $Q_n(g)$ of $g$. The following procedure using well-known
observations \cite{GB} shows how $Q_n$ helps to find the numbers $p$ and $q$.

\begin{itemize}
\item[]{\bf Step 1.} Randomly choose $x\in\ZZ_n^*$.
\vspace{2mm}

\item[]{\bf Step 2.} Set $y=Q_n(x^m)$. If $x=y$, then go to Step 1.
\vspace{2mm}

\item[]{\bf Step 3.} Output $q=\GCD(x-y,n)$ and $p=n/q$.
\end{itemize}
\vspace{2mm}

\noindent Let $x=x_p\times x_q$ and $y=y_p\times y_q$ where
$x_p,y_p\in\ZZ^*_p$ and $x_q,y_q\in\ZZ^*_q$. From Step~2 it follows that
$x_q^m=y_q^m$. On the other hand, since $n\in D_{N,m}$, we have
$\GCD(q-1,m)=1$. Thus $x_q=y_q\,(\mod q)$ and hence
$$
x=x_q=y_q=y\ \,(\mod q).
$$
So, $x-y\ne 0\ \,(\mod n)$ is a multiple of $q$. To complete the proof we
note that since $m=O(1)$, the loop of Steps 1,2 terminates with a large
probability after a polynomial number of iterations.\bull

Unfortunately, we don't know how to apply this lemma without assumption (*)
because in our case the system $\S_{N,m,n}$ includes the set $R\in R_{m,n}$.
However, from it we obtain the following statement.

\thrml{th2}
Under assumption (*) the triple $\S_{N,m,n}$ for an odd prime $m$ is a
homomorphic cryptosystem over $\ZZ_m^+$; in particular, the class
$\G_{crypt}$ contains each cyclic group of a prime order.\bull
\ethrm

We complete the section by mentioning that $\S_{N,m,n}$ can be slightly modified to
avoid the applying of Rabin's algorithm for finding roots of polynomials over finite
fields. In principle, to implement the decryption algorithm it suffices to determine
whether a given number $g\in G$ belongs to the group $G_0$ or not. However, this can
be done by observing that $g\in G_0$ iff $g_p^{(p-1)/m}=1\,(\mod p)$ where
$g_p$ is the component of~$g$ in the factor $\ZZ_p^*$ of
$G=\ZZ_p^*\times \ZZ_q^*$.

\section{Homomorphic cryptosystems using free products}
\label{frpr}

\subs
Throughout the section for a set $X$ we denote by $W(X)$ the set of all
words in the alphabet~$X$. For an element $w\in W(X)$ we denote by $|w|$
the length of~$w$.

Let $G_1,\ldots,G_m$ be a set of $m\ge 1$ pairwise disjoint finite groups.
For $i=1,\ldots,m$ set
$$
X_i=G_i\setminus\{1_{G_i}\},\quad \R_i=\{xyz\in W(X_i):\ x,y,z\in X_i,\ z^{-1}=xy\}.
$$
Then $G_i=\lg X_i; \R_i\rg$, i.e. $G_i$ is the group given by the set~$X_i$ of
generators and the set $\R_i$ of relations. Set $X_G=\cup_{i=1}^m X_i$ and
$\R_G=\cup_{i=1}^m\R_i$. The group
$$
G=G_1*\cdots*G_m=\lg X_G;\ \R_G\rg
$$
is called the {\it free product} of the groups~$G_1,\ldots,G_m$
(see~\cite{MCS}). From the definition it
follows that each element of $G$ can be represented by the uniquely
determined (canonical) word of $W_G=W(X_G)$ such that no two adjacent letters of it belong to the same
set among the sets~$X_i$. This enables us to identify $G$ with the subset
of~$W_G$ consisting of all such words. Thus $G=\{\ov w\in W_G:\ w\in W_G\}$
where $\ov w$ is the canonical word corresponding to a word~$w$. In
particular, $\ov g=g$ for all $g\in G$. Due to identifying the groups
$G_1,\ldots,G_m$ and $G$ with the corresponding subsets of the set~$W_G$,
we will assume below that the identities of these groups are equal to
the empty word of~$W_G$.

Suppose we are given epimorphisms $f_i:G_i\to K_i$, $i=1,\ldots,m$. Assuming
the groups $K_1,\ldots,K_m$ being pairwise disjoint we set $K=K_1*\cdots*K_m$
and $W_K=W(X_K)$ where $X_K=\cup_{i=1}^m(K_i\setminus\{1_{K_i}\})$. Then the
natural surjection $W_G\to W_K$ replacing the elements of~$X_G$ by their
images in $X_K$ with respect to corresponding~$f_i$, induces an epimorphism
\qtnl{l1}
f^*:G\to K,\quad f^*|_{X_i}=f_i,\quad i=1,\ldots,m.
\eqtn
Moreover, since the conditions $f^*|_{X_i}=f_i$ define the images of the
generators of~$G$, the epimorphism $f^*$ is the unique epimorphism from $G$
onto $K$ satisfying these conditions.

\subsn
\label{ssl3}
Let us study the kernel of the epimorphism~$f^*$. To do this suppose that
$K_i$ is a cyclic group of a prime order and $(R_i,A_i,P_i)$ is a homomorphic
cryptosystem over the group $K_i$ with respect to the epimorphism
$f_i:G_i\to K_i$, described in Section~\ref{expc}, $i=1,\ldots,m$.
Let $w_1,w_2\in W_G$ and $x\in G_i$ for some~$i$. By an {\it elementary
transformation} of the word $w_1xw_2\in W_G$ we mean replacing of~$x$ by
a certain word $x_1P_j(a_j)x_2$ where $x_1,x_2\in G_i$ with
$x=\ov{x_1x_2}$ and $a_j\in A_j$ for some $j$:
\qtnl{spl2}
w_1xw_2\quad\to\quad w_1(x_1P_j(a_j)x_2)w_2.
\eqtn
Denote by $W_0$ the set of all words of $W_G$ which can be obtained from the
empty word by a sequence of elementary transformations and set
$G_0=\{\ov w:\ w\in W_0\}$.

\lmml{spl1}
In the above notations, $\ker(f^*)=G_0$.
\elmm
\proof First, let $w=w_1xw_2$ where $w_1,w_2\in W_G$ and $x\in G_i$ for
some~$i$, to be a word of $W_G$. Suppose that a word $w'\in W_G$ is obtained
from~$w$ by the elementary transformation~(\ref{spl2}). Then the words
$\ov w,\ov{w'}$ belong or not to the group $\ker(f^*)$ simultaneously.
Indeed,
$$
f^*(\ov w')=f^*(\ov w_1\ov x_1\ov{P_j(a_j)}\ov x_2\ov w_2)=
f^*(\ov w_1)f^*(\ov x_1)f^*(\ov x_2)f^*(\ov w_2)=
$$
\qtnl{abc1}
f^*(\ov w_1)f^*(\ov x)f^*(\ov w_2)=f^*(\ov{w_1xw_2})=f^*(\ov w).
\eqtn
So, the inclusion $\ker(f^*)\supset G_0$ follows by the induction on the number
of elementary transformation used for constructing an element of~$G_0$.

Conversely, let $w'\in\ker(f^*)$. Let us prove that $w'\in G_0$ by the
induction on~$|w'|$. If $|w'|=0$, then the statement is obvious. Suppose
$|w'|>0$. Then $w'=w_1xw_2$ for some $w_1,w_2\in W_G$ and
$x\in\ker(f_i)$ for some~$i$. (Indeed, otherwise since $\ov{w'}=w'$, we
conclude that $|w'|=|f^*(w')|$. So, $|f^*(w')|>0$ which contradicts the fact
that $w'\in\ker(f^*)$.) So, $w'$ is obtained from the word $w=w_1w_2$ of $W_G$
by the elementary transformation~(\ref{spl2}).
Since $w'\in\ker(f^*)$ from (\ref{abc1}) it follows that
$\ov w\in\ker(f^*)$. On the other hand, it is easy to see that
$|\ov w|\le|w|<|w'|$. So, by the induction hypothesis we conclude that
$\ov w\in G_0$. By the definition of $\ov w$ this implies that $w'\in G_0$.
Thus $\ker(f^*)\subset G_0$ and we are done.\bull

Let $g\in\ker(f^*)$. Then from Lemma~\ref{spl1} it follows that $g$ can be
obtained from the empty word by a sequence of elementary transformations.
Moreover, the proof of this lemma implies that there exists such a sequence
consisting of at most $|g|$ elementary transformations. Any such sequence is
called a {\it proof} for $g$ (more precisely, a proof of the membership of
$g\in\ker(f^*)$, cf.~(H3) in Definition~\ref{md}). It is easy to see that
any elementary transformation~(\ref{spl2}) is uniquely determined by the
following data: the position of the letter~$x\in G_i$, the word
$x_1x_2x^{-1}\in\R_j$ and the element $a_j\in A_j$. Thus any proof for
the element $g$ can be represented by a word $p$ in the alphabet
$\NN\times\R_G\times(\cup_iA_i)$. One can see that in this case $|p|$
is bounded by a polynomial in~$|g|$.

We define $A^*$ to be the set of all proofs for the elements of~$\ker(f^*)$.
It should be stressed that $A^*$ includes only ``short'' (consisting of
at most $|g|$ elementary transformations) proofs for an element
$g\in\ker(f^*)$ and does not contain ``proofs'' for all words of~$W_0$.
For a given~$s$ one can generate a random element of $A^*$ of the length~$s$
in time $s^{O(1)}$. Indeed, due to the definition of the elementary
transformation~(\ref{spl2}) it suffices to choose randomly positions in
a current word and elements of $A_i$ for all $i=1,\ldots,m$. However, this
can be done with the help of the algorithms of the homomorphic cryptosystem
$(R_i,A_i,P_i)$ over~$K_i$ (see condition~(H1) of Definition~\ref{md}).

\lmml{spl3}
The image of the mapping $P^*:A^*\to G$ defined by $P^*(a)=g$ iff $a$ is a
proof for~$g$, equals $\ker(f^*)$. Moreover, the following statements hold:
\nmrt
\tm{i1} given $a\in A^*$ the element $P^*(a)$ can be found in polynomial
time in $|a|$,
\tm{i2} if for each $i\in\{1,\ldots,m\}$ there is an oracle~$Q_i$ which for
any element
$g_i\in\ker(f_i)$ produces a certain $a_i\in P_i^{-1}(g_i)$, then given
$g\in\ker(f^*)$ a proof $a\in A^*$ for $g$ can be found by means of at most
$|g|^2$ calls of oracles~$Q_i$ for $P_i^{-1}(g_i)$, $i=1,\ldots,m$,
$g_i\in\ker(f_i)$,
\tm{i3} for each $i\in\{1,\ldots,m\}$ and $g\in\ker(f_i)$ the problem of
finding an element in $P_i^{-1}(g)$ is polynomial time reducible to the
problem of finding an element in $(P^*)^{-1}(g)$.
\enmrt
\elmm
\proof The equality $\im(P^*)=\ker(f^*)$ follows from Lemma~\ref{spl1}.
Statement (i1) follows from the fact that any elementary transformation
(\ref{spl2}) of $g\in W_G$ is reduced to finding $P_j(a_j)$ which can be done in
polynomial in $|a_j|\le|a|$ time. To prove statement (i2) one can apply
the following obvious procedure testing membership of a word $w\in W_G$ to
the set~$W_0$.

\begin{itemize}
\item[]{\bf Step 1.} Using multiplications in the groups $G_i$, $i=1,\ldots,m$,
find the canonical word $\ov w$ of the word~$w$.
\vspace{2mm}

\item[]{\bf Step 2.} Using the oracles $Q_i$, $i=1,\ldots,m$,
delete any letter $x\in\ker(f_i)$ from the (current) word~$\ov w$. If there
was at least one deletion, then go to Step~1.
\vspace{2mm}

\item[]{\bf Step 3.} If the resulting word is empty, then $w\in W_0$.
\end{itemize}
\vspace{2mm}

In fact, this procedure is the algorithmic version of the proof of
Lemma~\ref{spl1}. To find a proof for arbitrary $g\in\ker(f^*)$, it suffices
to apply the above procedure to the word $g\in W_G$ and to collect all
results at Steps~1 and~2. Since the number of them is at most $|g|$, and
the number of calls the oracles $Q_i$ at Step~2 is also at most $|g|$,
statement~(i2) follows.

To prove statement~(i3) let $i\in\{1,\ldots,m\}$ and $g\in G_i$. Then since
obviously $g\in\ker(f_i)$ iff $g\in\ker(f^*)$, one can test whether
$g\in\ker(f_i)$  by means of an algorithm finding $(P^*)^{-1}$. Moreover, if
$g\in\ker(f_i)$, then this algorithm yields a proof from $A^*$ for the
element~$g$. Set $T$ to be the set of all elements $a_j\in A_i$
of elementary transformations~(\ref{spl2}) belonging to this proof.
Then $g=\prod_{a_j\in T}P(a_j)$. Since the set $A_i$ is an
Abelian group and the mapping $P_i:A_i\to G_i$ is a homomorphism, this implies that
$a=\prod_{a_j\in T}a_j$ is a proof for~$g$ and we are done.\bull

Let us describe a special system of distinct representatives of the right
cosets of $G$ with respect to $\ker(f^*)$. Set
$W_R=W(\cup_iR_i)$. Then $W_R\subset W_G$ and the set
\qtnl{orss}
R^*=G\cap W_R
\eqtn
is a system of distinct representatives of the right cosets of $G$ with
respect to $\ker(f^*)$. Indeed, $R^*$ consists of all the words of~$W_R$ which
are canonical words of $W_G$. So, no two elements of $R^*$ belong to
the same right coset of $G$ with respect to $\ker(f^*)$. Thus our claim
follows from the fact that the restriction of the mapping $f^*$ on $R^*$ is
the bijection from $R^*$ onto $K$ coinciding with $f_i$ on $R_i$,
$i=1,\ldots,m$.

\subsn\label{kkko}
We need one more special homomorphism of a free product. To do this we
recall some facts on semidirect products of groups (see e.g.~\cite{KM}).
Suppose that $K_1,K_2$ are groups and $\varphi:K_2\to\aut(K_1)$ is a
homomorphism. Then the set $K_1\times K_2$ forms a group with the
multiplication given by
$$
(k_1,k_2)(l_1,l_2)=(k_1(l_1)^{\varphi(k_2^{-1})},k_2l_2),\qquad
k_1,l_1\in K_1,\quad k_2,l_2\in K_2.
$$
This group is called a {\it semidirect product} of $K_1$ and $K_2$ (with
respect to the homomorphism~$\varphi$) and is denoted by $H=\Pi(K_1,K_2)$.
One can see that it contains the subgroup $K'_i$ (isomorphic to $K_i$)
consisting of all pairs with $1_{K_{3-i}}$ as the $(3-i)$-th coordinate, $i=1,2$. Moreover, $K'_1$
is a normal subgroup of $H$, $K'_1\cap K'_2=\{1_H\}$ and $H=K'_1K'_2$.
In general, if an arbitrary group~$H$ have such two subgroups $K'_1,K'_2$,
then it is isomorphic to the semidirect product of them with respect to the
homomorphism $\varphi$ induced by the action of $K'_2$ on $K'_1$ by the
conjugation. In what follows we shall identify $K_i$ with $K'_i$, $i=1,2$.
We also extend the definition of the semidirect product to arbitrary number
of factors by means of setting for $m\ge 3$
$$
\Pi(K_1,K_2,\ldots,K_m)=\Pi(K_1,\Pi(K_2,\ldots,K_m))
$$
with respect to the suitable homomorphisms $\varphi$. Thus
$\Pi(K_i,\ldots,K_m)=\Pi(K_i,K^{(i)})$ where
$K^{(i)}=\Pi(K_{i+1},\ldots,K_m)$, for
all $i=1,\ldots,m-1$ (for $i=m-1$ we adopt that $\Pi(K_m)=K_m$).
In what follows the group $\Pi(K_1,K_2,\ldots,K_m)$
will be ``small'' and presented by its multiplication table. Thus, its
subgroups $K_1,\ldots,K_m$ are also small and for a given $i$ the homomorphism
$\varphi_i:K^{(i)}\to\aut(K_i)$ can be presented by indicating the permutations
$\varphi_i(k)$ of the set $K_i$ for all $k\in K^{(i)}$.

\lmml{ortd}
Let $H=\Pi(K_1,\ldots,K_m)$ and $K=K_1*\cdots*K_m$ for a set of pairwise
disjoint finite groups $K_1,\ldots,K_m$. Then there exists an epimorphism
$Q:K\to H$ such that given $k\in K$ one can find the element $Q(k)$
in time polynomial in $|k|$ and $|H|$.
\elmm
\proof Due to our assumptions we see that the set $H$ as well as the set
$$
\R_H=\{x^{-1}yxy'\in W_K:\ x\in K^{(i)},\ y\in K_i,\ y'=(x^{-1}yx)^{-1},\
i=1,\ldots,m-1\}
$$
are the subsets of the group~$K$. From the definition of the free product
it follows that the elements of $H$ are distinct elements of the
quotient group $\ov K$
obtained from $K$ by imposing the set $\R_H$ of relations. On the other hand,
from the definition of $\R_H$ it follows that any element $\ov k\in\ov K$
can be represented by an element of $K$ of the form $k_1\cdots k_m$
for some $k_i\in K_i$, $i=1,\ldots,m$, moreover this representation is unique,
since otherwise the equality of two such representations one could deduce
from the relations $\R_H$ which hold in the group~$H$ as well, but in the
group~$H$ any two such representations differ. After identifying $\ov K$ with the set
of all such elements, we see that the mapping
$$
\ov K\to H,\quad \ov k\mapsto k_1\cdots k_m
$$
is an isomorphism. Denote by $Q$ the composition of the natural
epimorpism $K\to\ov K$ with this isomorphism. Then the mapping $Q:K\to H$ is
an epimorphism and given $k\in K$ the computation of $Q(k)$ consists in
a reduction of $k$ modulo the relations of~$\R_H$ to the form $k_1\cdots k_m$.
This can be done by means of the following procedure

\begin{itemize}
\item[]{\bf Step 1.} If $k$ is the empty word, then output $Q(k)=k$.
\vspace{2mm}

\item[]{\bf Step 2.} Using the relations $x^{-1}yxy'\in\R_H$ with arbitrary
$x\in K^{(1)}$ and $y\in K_1$ reduce $k$ to the form $k_1\wk$ where
$k_1\in K_1$ and $\wk\in\wK$ with $\wK=K_2*\cdots*K_m$.
\vspace{2mm}

\item[]{\bf Step 3.} Applying the procedure recursively to $\wk\in\wK$
and $\wH=K^{(2)}$, output $Q(k)=k_1Q(\wk)$.
\end{itemize}
\vspace{2mm}

First, we observe that the length of any intermediate word in the above
procedure is at most $|k|$. Next, the number of recursive calls (at Step~3)
is at most $|m|$. Thus the procedure can be done in time polynomial in $|k|$
and $|H|$. Lemma is proved.\bull

\subsn
We are ready to describe the main construction of this section. Let
$K_1,\ldots,K_m$ be a set of pairwise disjoint cyclic groups of prime
orders and $H=\Pi(K_1,\ldots,K_m)$. Suppose that we are given a homorphic
cryptosystem
$(R_i,A_i,P_i)$ over the group $K_i$ with respect to the homomorphism
$f_i:G_i\to K_i$ from Section~\ref{expc}, $i=1,\ldots,m$. Without loss of
generality we assume that the groups $G_i$ are pairwise disjoint.
Set $G=G_1*\cdots*G_m$. Then from the definition of $f^*$ (see
formula~(\ref{l1})) and Lemma~\ref{ortd} it follows that the mapping
$f=Qf^*$ from $G$ to $H$ is an epimorphism.

\thrml{iuyt}
The triple $(R,A,P)$ where $R$ is an arbitrary set of distinct
representatives of the right cosets of $G$ with respect to $\ker(f)$,
provided that $R$ fulfils the condition (H2) of Definition~\ref{md},
$$
A=\{(a,r)\in A^*\times R^*:\ f(r)=1_H\},\quad
P:A\to G,\ (a,r)\mapsto \ov{P^*(a)r}
$$
with $A^*$, $P^*$ defined in Subsection~\ref{ssl3} and $R^*$ defined
in~(\ref{orss}), is a homomorphic cryptosystem over the group~$H$ with
respect to the homomorphism $f:G\to H$.
\ethrm
\proof From the definition of $R^*$ it follows that given $g\in G$ there exist
uniquely determined $g_0\in\ker(f^*)$ and $r\in R^*$ such that $g=g_0r$. So,
$f(g)=Q(f^*(g_0)f^*(r))=Q(f^*(r))=f(r)$. Thus $g\in\ker(f)$ iff $f(r)=1$.
By Lemma~\ref{spl3} this implies that $\im(P)=\ker(f)$.
To check the condition (H1) of Definition~\ref{md} we have
to show how to get random elements of~$A$. This follows from the remarks before
Lemma~\ref{spl3} for random generating elements of $A^*$, whereas the sets
$R_i$, $i=1,\ldots,m$, are given explicitly. Thus it remains to verify that
$P$ is a trapdoor function (i.e. the condition (H3)).

First, we observe that by statement~(i1) of Lemma~\ref{spl3} and by
Lemma~\ref{ortd} the mapping $P$ is polynomial time computable. Second,
by condition (H3) for homomorphic cryptosystems $(R_i,A_i,P_i)$ there
exists an algorithm that given $i\in\{1,\ldots,m\}$ and $g_i\in G_i$
efficiently finds an element of the set $P_i^{-1}(g_i)$. Let us show that
it suffices to invert~$P$. Indeed, in this case given $g\in G$ the element
$f^*(g)\in K$ can be found efficiently. Since $g=g_0r$ for uniquely
determined $g_0\in\ker(f^*)$ and $r\in R^*$, and $f^*(r)=f^*(g)$, one can
compute the element $r$ and hence the element $g_0=gr^{-1}$ within the same
time. By statement~(i2) of Lemma~\ref{spl3} we can also find an element
$a\in A^*$ such that $P^*(a)=g_0$. Thus to invert $P$ it suffices to
test whether $f(r)=1_H$ holds (if $f(r)=1_H$, then $(a,r)\in A$ and
$P(a,r)=g$). We have $f(r)=Q(f^*(r))$. Next, by condition (H2) for
homomorphic cryptosystems $(R_i,A_i,P_i)$ we can find the element $f^*(r)$
and so by Lemma~\ref{ortd} the element $Q(f^*(r))$, and finally
test the equality $f(r)=1_H$.

Suppose that one can invert $P$ efficiently. Let $g\in G$. If
$g\not\in\ker(f)$, then obviously $g\not\in\ker(f^*)$.
Let now $g\in\ker(f)$ and $(a,r)\in A$ be a proof for~$g$, i.e.
$\ov{P^*(a)r}=g$. Since $r$ belongs to the right transversal $R^*$
of $\ker(f^*)$ in $G$, it follows that $g\in\ker(f^*)$ iff $r=1_G$.
Moreover, if $r=1_G$, then obviously $P^*(a)=g$. Thus the problem of
finding $(P^*)^{-1}$ is polynomial time reducible to the problem of
finding $P^{-1}$. So by statement (i3) of Lemma~\ref{spl3} the problem of
finding $P_i^{-1}$, $i=1,\ldots,m$, is polynomial time reducible to the
problem of finding $P^{-1}$. Thus $P$ is a trapdoor function which
completes the proof.\bull

Observe that one can explicitly produce a set $R$ satisfying the condition
of Theorem~\ref{iuyt} (i.e. the condition (H2)). Namely, for each element
$h\in H$ find a representation $h=k_1\cdots k_m$ where $k_i\in K_i$ (cf.
Subsection~\ref{kkko}), and take $r_i\in R_i$ such that $f_i(r_i)=k_i$,
$i=1,\ldots,m$. Then the set of all elements $r_1\cdots r_m$ for all
$h\in H$ can be chosen as the set~$R$.

{From} Theorems~\ref{th2} and~\ref{iuyt} we immediately obtain the following
statement.

\crllrl{l10}
Let $K_1,\dots,K_m$ be a set of pairwise disjoint cyclic groups of prime
orders, $m\ge 1$. Then $\Pi(K_1,\ldots,K_m)\in\G_{crypt}$. \bull
\ecrllr

\subsn
The special cases of a semidirect product are the direct and wreath products.
Indeed, in the latter case the resulting group is a semidirect product of
the direct power of the first group (with the number of the factors being
equal to the order of the second group) by the second group which acts on
the product by permutations of direct factors, see e.g.~\cite{KM}. Thus as
an immediate consequence of Theorem~\ref{iuyt} we conclude that the class
$\G_{crypt}$ contains direct and wreath products of cyclic
groups of prime orders (cf. Corollary~\ref{l10}). Using this fact we can
prove the main result of the paper.

\thrml{th3}
Any solvable nonidentity group belongs to the class $\G_{crypt}$.
\ethrm
\proof It is a well-known fact that any solvable group can be constructed from
a cyclic group of prime order by a sequence of cyclic extensions. On the
other hand, from~\cite[Theorem 6.2.8]{KM} it follows that any extension of
one group by another one is isomorphic to a subgroup of the wreath
product of them. So it suffices to verify (cf. Corollary~\ref{l10}) that
any nonidentity subgroup of a semidirect product of cyclic groups of
prime orders belongs to the class $\G_{crypt}$.

To do this let $H\in\G_{crypt}$ be such a group. Then there exists a homomorphic cryptosystem
$\S=(R,A,P)$ over $H$ with respect to some epimorphism $f:G\to H$. Without
loss of generality we assume that $\S$ is the homomorphic cryptosystem
from Theorem~\ref{iuyt}. Given explicitely a non-identity subgroup $H'$ of
$H$ set $G'=f^{-1}(H')$, $f'=f|_{G'}$ and $\S'=(R',A',P')$ where
$R'=R\cap f^{-1}(H')$,
$$
A'=\{(a,r)\in A^*\times (R^*)':\ f(r)=1_H\},\quad
P':A'\to G',\ (a,r)\mapsto \ov{P^*(a)r}
$$
and $(R^*)'=\{r\in R^*:\ f(r)\in H'\}$. Then $\S'$ is a homomorphic
cryptosystem over $H'$ with respect to the homomorphism $f':G'\to H'$.
Indeed, we present the group $G'$ as a subgroup of $G$
generated by the sets $\im(P')$ and $R'$. (In this presentation of $G'$ we
would be unable to recognize its elements in $G$, but we do not need this.)
Now the first two conditions of the Definition~\ref{md} are satisfied for
$\S'$ because they are satisfied for $\S$ (to generate a random element
of $A'$, it suffices to generate a random element $r'$ of
$(R^*)'$ and for this purpose one can generate a random $r\in R^*$ and set
$r'=\ov{r\wt r}$ where
$\wt r$ is the element of $R'$ such that $f(r)f(\wt r)^{-1}\in H'$). Since
$\ker(f')=\ker(f)$, we have $\im(P')=\ker(f')$ and condition (H3)
is also satisfied for $\S'$ because it is satisfied for~$\S$.\bull

It should be remarked that the construction of a homomorphic cryptosystem
over a solvable group~$H$ described in this section is rather theoretical.
The computational complexity of the underlying algorithms is bounded by a
polynomial the degree of which is a function of~$|H|$. Besides, the size of
representing $G$ could be exponential in $|H|$ due to involving wreath
products. However, it seems that more careful implementation can be developed
for small groups.

{From} Theorem~\ref{th3} it follows that there exists a homomorphic
cryptosystem over the group $\sym(n)$ for $n\le 4$. It would be interesting
to construct a homomorphic cryptosystem over an arbitrary symmetric group,
because such a system would provide secret computations with any permutation
group and moreover an implementation of any boolean circuit in the sense
of~\cite{Ba}. In this connection we remark that every boolean circuit of
logarithmic depth can be implemented by a polynomial-time computation in an
arbitrary nonsolvable group (see \cite{Ba}). On the other hand, it was
proved in~\cite{Ba}
that over an arbitrary nilpotent group not any boolean circuit can be
implemented. If the group is not nilpotent but is solvable, then only an
exponential size implementation is known~\cite{Ba} and it is conjectured that
one is unable to do better. Thus, if the latter conjecture was wrong, then
combining with Theorem~\ref{th3} would enable us to encrypt any boolean
circuit.

\vspace{5mm}

{\bf Acknowledgements.}
The authors would like to thank the Max-Planck Institut fuer Mathematik (Bonn)
during the stay in which this paper was done; also Eberhard Becker who has
drawn the authors attention to homomorphic cryptosystems and Igor Shparlinski
for useful discussions.

\end{document}